\documentclass[preprint,aps,12pt,showpacs,nofootinbib,tightenlines]{revtex4-1}
\usepackage{amsmath}
\usepackage{amssymb}
\usepackage{epsfig}
\usepackage{graphicx}
\begin{document}

\newcommand{\beq}{\begin{eqnarray}}
\newcommand{\eeq}{\end{eqnarray}}

\newcommand{\non}{\nonumber\\ }
\newcommand{\ov}{\overline}
\newcommand{\psl}{ p\hspace{-1.6truemm}/ }
\newcommand{\nsl}{ n\hspace{-2.2truemm}/ }
\newcommand{\vsl}{ v\hspace{-2.2truemm}/ }
\newcommand{\etar}{ \eta^\prime }
\newcommand{\etap}{ \eta^{(\prime)} }
\newcommand{\jpsi}{ J/\Psi }

%%%%%%%%%%%%%%%%%%%
\def \ctp{{\bf Commun.Theor.Phys. } }
\def \epjc{{\bf Eur.Phys.J. C} }
\def \ijmpa{{\bf Int.J.Mod.Phys. A} }
\def \jhep{ {\bf JHEP  } }
\def \jpg{ {\bf J.Phys. G} }
\def \npb{ {\bf Nucl.Phys. B} }
\def \plb{ {\bf Phys.Lett. B} }
\def \ppnp{ {\bf Prog.Part.$\&$ Nucl.Phys} }
\def \prd{ {\bf Phys.Rev. D} }
\def \prl{ {\bf Phys.Rev.Lett.}  }
\def \ptp{ {\bf Prog. Theor. Phys. }  }
\def \rmp{ {\bf Rev.Mod.Phys. }  }
\def \zpc{ {\bf Z.Phys.C}  }
%%%%%%%%%%%%%%%%%%%%
%%%%%%%%%%%%%%%%%%%%%%%%%%%%%%%%%%%%%%%%%%%%%%%%
\title{Semileptonic decays $B/B_s \to (\eta,\etar, G)(l^+l^-,l\bar{\nu},\nu\bar{\nu} )$
in the perturbative QCD approach beyond the leading order}
\author{Wen-Fei Wang, Ying-Ying Fan, Min Liu, and Zhen-Jun Xiao
\footnote{Email Address: xiaozhenjun@njnu.edu.cn} }
\affiliation{ Department of Physics and Institute of Theoretical Physics,\\
Nanjing Normal University, Nanjing, Jiangsu 210023, People's Republic of China}
\date{\today}
%---------------------------------------------------------------------------------------------------------------%
\begin{abstract}
In this paper we make a systematic study of the semileptonic decays
$B/B_s \to (\eta,\etar, G)(l^+l^-,l\bar{\nu},\nu\bar{\nu} )$
by employing the perturbative QCD (pQCD) factorization approach.
The next-to-leading-order (NLO) contributions to the relevant
form factors are included, and the ordinary $\eta$-$\etar$ mixing scheme
and the $\eta$-$\etar$-$G$ mixing scheme are considered separately,
where $G$ denotes a pseudoscalar glueball. The numerical results and the
phenomenological analysis indicate that
(a) the NLO contributions to the relevant form factors
provide $25\%$ enhancement to the leading-order pQCD predictions for
the branching ratios $Br(B^-\to \eta^{(')} l^-\bar{\nu}_l)$, leading to
a good agreement between the predictions and the data;
(b) for all considered decays, the pQCD results are basically
consistent with those from other different theoretical models;
(c) the pQCD predictions in the two considered mixing schemes agree
well with each other within theoretical errors. The outcomes
presented here can be tested by LHCb and forthcoming Super-$B$ experiments.
\end{abstract}


\maketitle
%---------------------------------------------------------------------------------------------------------------%

The pseudoscalar mesons $\eta$ and $\etar$ are rather different from other light pseudoscalar
mesons $\pi$ or $K$, not only for the large mass of the $\etar$ meson, but also
for their mixing and possible gluonic components. These features caught more attention
recently because of the so-called $B\to K \etap$ puzzle. The two-body charmless hadronic $B$ meson
decays involving the $\etap$ final states and several different $\eta$-$\etar$
mixing schemes have been investigated intensively in the standard model (SM)
(see for instance \cite{lipkin91,yy01,bn651,kou02,zjx2008,fan2012})
and in the new physics models beyond the SM (see for instance \cite{kagan,xiao99}).
The semileptonic $B/B_s$ meson decays, studied in
Refs.~\cite{kim2001,kim-plb590-223,prd-75-054003,chen07,choi10,chen10,wu06,azizi10,semil-2012a},
also play an important role in improving our understanding about the nature of the $\eta$
and $\etar$ mesons.

Many precision measurements of the branching ratios and
$CP$ asymmetries for relevant $B$ meson decays,
such as $B\to K \etap$ and $B_s \to \jpsi \etap$, have become available \cite{pdg2012},
and been interpreted successfully within the SM
in Refs.~\cite{fan2012,liu-86-011501} by employing the perturbative QCD (pQCD)
factorization approach \cite{li2003}.
As for the $B_{d,s}\to ( \eta,\etar,G)(l^+l^-,l\bar{\nu},\nu\bar{\nu} )$
semileptonic decays, which will be studied in this paper,
only two of them, $B^+\to \etap l^+\nu_l$, were observed so far
\cite{babar-83-052011,pdg2012}.
For the others, experimental data
are not yet available, and we have to wait for measurements
at LHCb or at the forthcoming super-$B$ factory.

On the theory side, the previous predictions given in
Refs.~\cite{kim2001,kim-plb590-223}, where the form factors from QCD
sum rules or lattice QCD were adopted as inputs, basically agree with
the measured values.
The $B$ meson semileptonic decays, on the other hand, are frequently used
to extract the corresponding $B\to (P,V)$ transition form factors,
with $P$ and $V$ denoting the light
pseudoscalar and vector mesons, respectively, such as $\pi, K, \rho, \cdots$
when relevant data are available.

Analogous to Ref.~\cite{prd86-114025}, we will make a
systematic pQCD study of the semileptonic decays
$B/B_s \to(\eta,\etar,G)(l^+l^-, l\bar\nu, \nu\bar\nu)$ here,
where $G$ stands for a physical pseudoscalar glueball, and compare our pQCD
predictions with existing calculations and numerical results.
Based on the assumption of the $SU(3)$ flavor symmetry, we will
extend the NLO pQCD calculation for the $B \to \pi$ form factors
in Ref.~\cite{li-85074004} to the $B_{(s)}\to \eta_{q,s}$ cases, $\eta_{q,s}$
being the flavor eigenstates of the light and strange quarks, respectively.
The $B_{(s)}\to \eta_g$ transitions form factors will be also calculated in
the pQCD factorization approach, where $\eta_g$ represents the unmixed pseudoscalar
glueball. The relevant Feynman diagrams for the $B_{(s)} \to\eta_{q,s}$ and
$B_{(s)} \to G$ transitions are displayed in Fig.1.

\begin{figure}[thb]
%\vspace{-0.5cm}
\begin{center}
\includegraphics[scale=0.6]{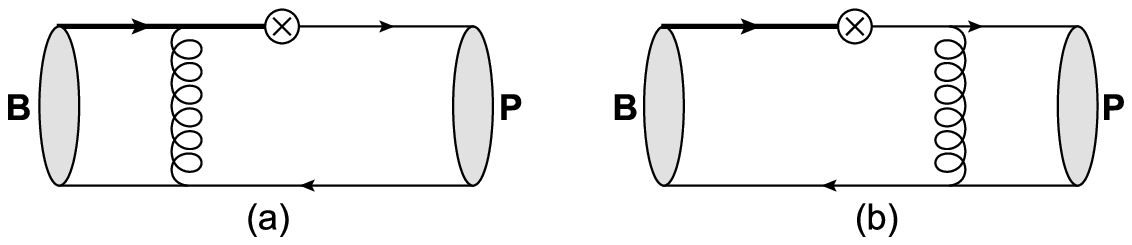}
\includegraphics[scale=0.8]{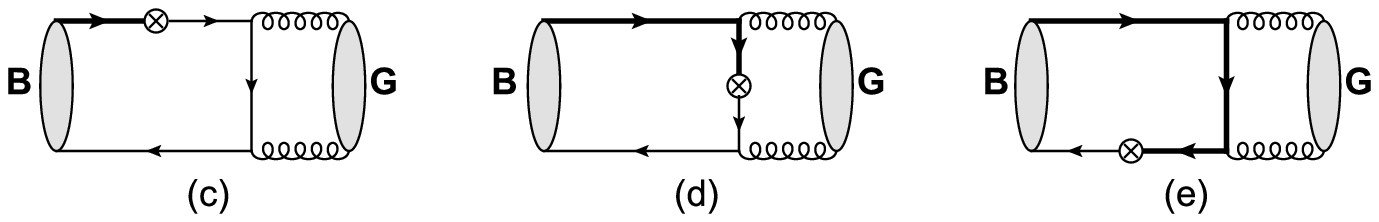}
\end{center}
\vspace{-0.5cm}
\caption{ Feynman diagrams: (a) and (b) for the $B_{(s)} \to \eta_{q,s}$ transitions, and
(c)-(e) for the $B_{(s)} \to G$ transitions. The symbol $\otimes$ refers to the weak
vertex where the final-state lepton pairs are emitted, $P$ and $G$ stand for
the $\eta_{q,s}$ mesons and the glueball, respectively.} \label{fig:fig1}
\end{figure}

We will consider two different meson mixing schemes in our
calculation, and compare the corresponding numerical results. The
first mixing scheme is the conventional Feldmann-Kroll-Stech (FKS)
$\eta$-$\etar$ mixing scheme \cite{fks98}, in which the physical states
$\eta$ and $\etar$ are written as
\beq \left(\begin{array}{c} \eta \\
\eta^{\prime}\end{array} \right) = \left(\begin{array}{cc}
 \cos\phi& -\sin\phi\\
\sin\phi& \cos\phi\\ \end{array} \right) \left(\begin{array}{c} \eta_q
\\ \eta_s\end{array} \right)\;,
\label{eq:e-ep2}
\eeq
with the flavor states $\eta_q=(u\bar{u}+d\bar{d})/\sqrt{2}$
and $\eta_s=s\bar{s}$, and the mixing angle $\phi$ \cite{fks98}.
The three input parameters $f_q, f_s$ and $\phi$ in the FKS mixing
scheme have been extracted from data of relevant exclusive
processes \cite{fks98}: $f_q=(1.07\pm 0.02)f_{\pi}$, $f_s=(1.34\pm
0.06)f_{\pi}$, and $\phi=39.3^\circ\pm 1.0^\circ$ with $f_\pi=0.13$
GeV.

In the second mixing scheme, i.e., the $\eta$-$\etar$-$G$ mixing scheme defined in
Ref.~\cite{li-79-014024}, the physical states $\eta$, $\etar$
and $G$ are related to $\eta_q$, $\eta_s$, and $\eta_g$ through the rotation
\beq
\label{qsg}
\left( \begin{array}{c}    |\eta\rangle \\ |\eta'\rangle\\|G\rangle   \end{array} \right)
   = U(\theta,\phi,\phi_G)  \left( \begin{array}{c}
    |\eta_q\rangle \\ |\eta_s\rangle\\|\eta_g\rangle  \end{array} \right) \;,
\eeq
where the mixing matrix $U(\theta,\phi,\phi_G)$ has been given in Eq.(4) of
Ref.~\cite{li-79-014024} with $\phi=\theta+54.7^\circ$ and $\phi_G\sim 30^\circ$.

In the numerical analysis, we assume the same functional form for
the $B$ and $B_s$ meson distribution amplitudes
we did in Ref. \cite{prd86-114025}.
For the $\eta_{q}$ and $\eta_{s}$ meson, we adopt the distribution amplitudes
given in Refs. \cite{pball-71-014015,wfun-pball-pi}. The relevant
Gegenbauer moments take the values
$ a_2^{\eta_{q(s)}}=0.115$, $a_4^{\eta_{q(s)}}=-0.015$, $\eta_3^{\eta_{q(s)}}=0.013$,
and $\omega_3^{\eta_{q(s)}} =-3$ \cite{pball-71-014015}.
The leading-twist distribution amplitude of the unmixed glueball $\eta_g$ is
defined by \cite{li-74-074024,ali-367}
\beq
\langle\eta_g(p_2)|A_{[\mu}^a(z)A_{\nu]}^b(0)|0\rangle
&=&\frac{f_{\eta_g} C_F  \delta^{ab}}{96}\epsilon_{\mu\nu\rho\sigma}\,
\frac{n^\rho p_2^\sigma\,}{n\cdot p_2} \int_0^1dx
e^{ix p_2 \cdot z}\frac{\phi^{G}(x)}{x(1-x)}\;,
\label{qgn}
\eeq
with the decay constant $f_{\eta_g}=f_s$ ($\sqrt{2}f_q$) for the
$B_s\to\eta_g$ ($B \to \eta_g$) transition. The function $\phi^{G}(x)$
is expressed as \cite{kroll-67}
\beq
\phi^{G}(x)=B_2^{q(s)}5 x^2(1-x)^2(2x-1)\;,
\label{phiG}
\eeq
where the coefficients $B_2^q$ and $B_2^s$ for the $B\to\eta_q$ and
$B_s\to\eta_s$ transitions, respectively,
take the value $B_2^q=B_2^s\equiv B_2=4.6\pm 2.5$ \cite{li-74-074024,ali-367}.

%---------------------------------------------------------------------------------------------------------------%

In the $B$ meson (for simplicity, $B$ denotes
both the $B$ and $B_s$ mesons here) rest frame, the momentum $p_1$ of the $B$ meson
and $p_2$ of the final-state light pseudoscalar meson are written as
$p_1=\frac{m_B}{\sqrt{2}}(1,1,0_{\rm T})$ and $p_2=\frac{m_B}{\sqrt{2}}\rho(0,1,0_{\rm T})$
with the energy fraction $\rho=1-q^2/m_B^2$, where the lepton pair momentum is
defined by $q=p_1-p_2$.
The light spectator momenta $k_1$ in the $B$ meson and $k_2$ in the final-state meson are
parameterized as $k_1 =(x_1,0,k_{1{\rm T}})\frac{m_B}{\sqrt{2}}$ and
$k_2=(0,x_2\rho,k_{2{\rm T}})\frac{m_B}{\sqrt{2}}$, respectively.

For the $B\to P$ transition, the relevant form factors $F_{0,+}(q^2)$
and $F_T(q^2)$ have been defined for example in
Ref.~\cite{npb592-3} with the relation $F_{0}(0)=F_+(0)$.
For convenience, one usually considers
the auxiliary form factors $f_1(q^2)$ and $f_2(q^2)$ defined via
\beq
\langle P(p_2)|\bar{b}(0)\gamma_{\mu}q(0)|B(p_1)\rangle
= f_1(q^2)p_{1\mu}+f_2(q^2)p_{2\mu},
\eeq
in terms of which $F_+(q^2)$ and $F_0(q^2)$ are written as
\beq
\label{eq:fpfz}
F_+(q^2)&=&\frac{1}{2}\left [f_1(q^2)+f_2(q^2) \right ],\non
F_0(q^2)&=&\frac{1}{2}f_1(q^2)\left [1+\frac{q^2}{m_B^2-m_P^2} \right ]
+\frac{1}{2} f_2(q^2)\left [1-\frac{q^2}{m_B^2-m_P^2} \right ].
\eeq

The authors in Ref.~\cite{li-85074004} derived the $k_{\rm T}$-dependent
NLO hard kernel $H$ for the $B \to \pi$ transition form factors. Here we quote
their results directly, and
extend the expressions to the $B_{(s)}\to \eta_{q(s)}$ transitions
under the assumption of $SU(3)$ flavor symmetry. The hard kernel
$H$ is given, at NLO, by~\cite{li-85074004}
\beq
H=H^{(0)}(\alpha_s)+ H^{(1)}(\alpha_s^2)
= \left [ 1+F(x_1,x_2,\mu,\mu_f,\eta,\zeta_1) \right ] H^{(0)}(\alpha_s),
\eeq
where the expression of the NLO factor $F(x_1,x_2,\mu,\mu_f,\eta,\zeta_1)$ can be
found in Eq.(3) of Ref. \cite{prd86-114025}.

Employing the pQCD factorization approach with the inclusion of the Sudakov
factors and the threshold resummation effects \cite{prd86-114025},
we obtain the form factors $f_{1,2}(q^2)$ and $F_T(q^2)$ for the considered decays.
For the $B_s\to \eta_s$ and $B\to \eta_q$ transitions, for instance, the form factor
$f_1^{B_s\to\eta_s}(q^2)$ is of the form
%--------------------------------f1 f2 FT----------%
\beq
f_1^{B_s\to\eta_s}(q^2)&=& \sqrt{2} f_1^{B\to\eta_q}(q^2)=
16\pi C_Fm_B^2\int dx_1 dx_2\int b_1 db_1 b_2
db_2 \phi_B(x_1,b_1)\non
&& \times \Bigl \{ [ r_0 \left ( \phi^p(x_2)-\phi^t(x_2) \right ) \cdot
h_1(x_1,x_2,b_1,b_2) - r_0 x_1 \rho m_B^2\phi^{\sigma}(x_2)\non
 &&\times h_2(x_1,x_2,b_1,b_2) ]
\cdot \alpha_s(t_1) \exp \left [-S_{B\eta}(t_1) \right ] \non
&& + \left [ x_1 \left ( \rho\phi^a(x_2)-2r_0\phi^p(x_2) \right )+
4 r_0 x_1\phi^p(x_2) \right ] \non
&& \times h_1(x_2,x_1,b_2,b_1)\cdot
\alpha_s (t_2)\exp\left [-S_{B\eta}(t_2) \right] \Bigr \},
\label{eq:f1q2}
\eeq
with $C_F=4/3$, $r_0=m_0^{\eta_s}/m_{B_s}$, $\rho=1-q^2/m^2_{B_s}$  
($r_0=m_0^{\eta_q}/m_{B}$, $\rho=1-q^2/m^2_B$) 
for the $B_s\to \eta_s$ ($B\to \eta_q$) transition. 
We choose $m_{\eta_q}=0.18\pm 0.08 {\rm GeV}$ and
$m_{\eta_s}=0.69 {\rm GeV}$ as in Ref.~\cite{prd-75-054003}.
The hard functions $h_{1,2}$, the scales $t_{1,2}$ and the Sudakov factors
$\exp[-S_{B\eta}(t)]$ are the same as in Refs.~\cite{prd86-114025,lu-jpg37}.
For the $B/B_s\to\eta_g$ transitions, we found similar expressions of form factors.

One should note that the form factors in
Eq.~(\ref{eq:f1q2}) represent the leading-order (LO) pQCD predictions.
To include the NLO corrections, the coupling constant $\alpha_s$ in
Eq.~(\ref{eq:f1q2}) is changed into
$\alpha_s \cdot F(x_1,x_2,\rho,\mu_f,\mu,\zeta_1)$,
with the NLO factor $F(x_1,x_2,\rho,\mu_f,\mu,\zeta_1)$ being given
in \cite{prd86-114025}.
For the $B_{(s)}\to\eta_g$ transitions, only the LO
expressions are available now.
The formulas for the differential $b\to ul^-\bar{\nu_l}$,
$b\to sl^+l^-$, and $b\to s\nu\bar\nu$ decay widths are provided in 
Ref.~\cite{prd86-114025}.

The following input parameters (the masses and decay constants are all
in units of GeV) \cite{fks98,liu-86-011501,pdg2012} are adopted in
the numerical analysis: $\Lambda_{\bar{MS}}^{(f=4)}=0.287$,
$f_B=0.21,  f_{B_s}=0.23, m_0^q=1.50$, $m_0^s=1.90,
m_{B}=5.279$, $m_{B_s^0}=5.3663, \tau_{B^\pm}=1.638\; ps$,
$\tau_{B^0}= 1.525\; ps, \tau_{B_s^0}=1.472\; ps$,
$m_\tau = 1.777, m_b=4.8, m_W=80.4$, $m_t=172, m_\eta=0.548$,
and $m_{\eta^\prime}=0.958$. For the relevant CKM matrix elements, we take
$|V_{tb}|=0.999$, $|V_{ts}|=0.0403$ and $|V_{td}/V_{ts}|=0.211$~\cite{pdg2012}.

As explained in Ref.~\cite{prd86-114025}, the pQCD predictions for the
form factors $F_{0,+,T}(q^2)$ are reliable only at small $q^2$,
such as the region $0\leq q^2 \leq m_\tau^2$.
To get the form factors at larger $q^2$, one has to make an
extrapolation from the lower $q^2$ region. Therefore,
we perform the pQCD calculations for the $B/B_s\to (\eta_q,\eta_s, \eta_g)$
transition form factors in the range $0\leq q^2 \leq m_\tau^2$, and then
apply the extrapolation using the pole model parametrization
\beq
 F_i(q^2)=\frac{F_i(0)}{1-a(q^2/m_B^2)+b(q^2/m_B^2)^2}\;,
\label{eq:pole-1}
\eeq
where $F_i$ denotes a function among $F_{0,+,T}$, and $a, b$ are
the constants to be determined by the fitting procedure.

\begin{table}[thb] %55555555555555555555555555555
\begin{center}
\caption{pQCD predictions for the form factors $F_{0,+,T}(0)$
in the $B\to\eta_q$ and $B_s \to\eta_s$, and $B/B_s \to\eta_g$  transitions
at LO and NLO. }
\label{tab-Bq-Bss}
\begin{tabular}{l|l| c| l|c} \hline\hline
\ \ \ & &$ F_i(0)$ &$ $ & $F_i(0)(10^{-2})$  \\
 \hline
$F^{B\to\eta_q}_{0}$& LO & $0.17\pm0.02\pm0.02\pm0.01$ &$F^{B\to\eta_g}_{0}$ &$0.14^{+0.02}_{-0.01}\pm0.01\pm0.08$ \\
                    & NLO & $0.19^{+0.03}_{-0.02}\pm0.02\pm0.01$ && \\ \hline
$F^{B\to\eta_q}_{+}$ & LO & $0.17\pm0.02\pm0.02\pm0.01$ &$F^{B\to\eta_g}_{+}$ &$0.14^{+0.02}_{-0.01}\pm0.01\pm0.08$ \\
                     &NLO & $0.19^{+0.03}_{-0.02}\pm0.02\pm0.01$ && \\ \hline
$F^{B\to\eta_q}_{T}$& LO & $0.15\pm0.02\pm0.01\pm0.01$&$F^{B\to\eta_g}_{T}$&$0.10\pm0.01\pm0.01\pm0.06  $\ \\
                    &NLO & $0.17\pm0.02\pm0.02\pm0.01$&&  \\ \hline
$F^{B_s\to\eta_s}_{0}$& LO & $0.27^{+0.04}_{-0.03}\pm0.04\pm0.01$  &$F^{B_s\to\eta_g}_{0}$&$0.11\pm0.01\pm0.01\pm0.06$\ \\
                      & NLO & $0.31^{+0.05}_{-0.04}\pm0.04\pm0.01$ && \\ \hline
$F^{B_s\to\eta_s}_{+}$& LO & $0.27^{+0.04}_{-0.03}\pm0.04\pm0.01$ &$F^{B_s\to\eta_g}_{+}$ &$0.11\pm0.01\pm0.01\pm0.06$\ \\
                      &NLO & $0.31^{+0.05}_{-0.04}\pm0.04\pm0.01$ && \\ \hline
$F^{B_s\to\eta_s}_{T}$& LO & $0.27\pm0.04\pm0.04\pm0.01$ &$F^{B_s\to\eta_g}_{T}$ &$0.08\pm0.01\pm0.01\pm0.04$\ \\
                      & NLO & $0.31^{+0.05}_{-0.04}\pm0.04\pm0.01$ && \\
\hline\hline
\end{tabular}
\end{center}
\end{table}

%%%%%%%%%%%%%%%%%%%%%%%%%%%%%%%%%%%%%%%%%%%%%%%%vvv
\begin{table}[thb]
\begin{center}
\caption{LO and NLO pQCD predictions for the branching ratios of the considered
decays in the FKS mixing scheme. The relevant data \cite{pdg2012} and
other theoretical predictions
\cite{kim2001,chen07,chen10,choi10,wu06,azizi10} are listed in last two columns. }
\label{tab-brn1}
{\small
\begin{tabular}{lcccccl}  \hline\hline
  Decay modes  & ${\rm LO}$ &  $ {\rm pQCD_{NLO}}$& {\rm Set-A} & {\rm Set-B}& ${\rm Others}$ & Data \cite{pdg2012}\\
\hline %%%%%%%%%%%%%
$Br(B^-\to\eta l^- \bar\nu_l)(10^{-4})$              &0.33 & $ 0.41^{+0.12}_{-0.09}\pm 0.08^{+0.04}_{-0.03}$& $0.33^{+0.12}_{-0.10} $ & $0.37^{+0.14}_{-0.11} $  &$0.43\pm 0.08$\; \cite{kim2001} &$0.39\pm 0.08$  \\
$Br(B^-\to\eta \tau^- \bar\nu_\tau)(10^{-4})$        &0.19 & $ 0.24^{+0.07+0.05}_{-0.05-0.04}\pm 0.02$      & $0.20^{+0.07}_{-0.05} $ & $0.23^{+0.08}_{-0.06} $  & $0.29^{+0.07}_{-0.06}$\; \cite{chen10}   & \\
$Br(B^-\to\eta^\prime l^- \bar\nu_l)(10^{-4})$       &0.16 & $ 0.20^{+0.06}_{-0.04}\pm 0.04\pm 0.02 $       & $0.16^{+0.06}_{-0.05} $ & $0.17^{+0.06}_{-0.05} $  & $0.21\pm 0.04$\; \cite{kim2001} &$0.23\pm 0.08$\\
$Br(B^-\to\eta^\prime \tau^- \bar\nu_\tau)(10^{-4})$ &0.08 & $ 0.10^{+0.03}_{-0.02}\pm 0.02\pm 0.01$        & $0.08^{+0.03}_{-0.02} $ & $0.09^{+0.03}_{-0.02} $  &$0.13^{+0.03}_{-0.02}$\; \cite{chen10}  &\\
%%%%%%%%%%%%%
$Br(\bar{B}^0\to\eta l^+ l^-)(10^{-8})$              &0.39 & $0.48^{+0.14+0.10+0.05}_{-0.10-0.09-0.04} $    & $0.39^{+0.14}_{-0.11} $ & $0.45^{+0.17}_{-0.13} $  & $0.6$\; \cite{chen07}&\\
$Br(\bar{B}^0\to\eta \tau^+ \tau^-)(10^{-9})$        &0.83 & $0.98 ^{+0.28+0.19+0.08}_{-0.20-0.18-0.06} $   & $0.80^{+0.28}_{-0.22} $ & $0.92^{+0.33}_{-0.26} $  &$1.1\pm 0.1$\; \cite{wu06}  &\\
$Br(\bar{B}^0\to\eta \nu\bar\nu)(10^{-9})$           &0.31 & $0.38^{+0.11+0.08+0.03}_{-0.08-0.07-0.03} $    & $0.31^{+0.11}_{-0.09} $ & $0.36^{+0.13}_{-0.11} $  & &\\
%%%%%%%%%%%%%
$Br(\bar{B}^0\to\eta^\prime l^+ l^-)(10^{-8})$       &0.18 & $0.24^{+0.07+0.05}_{-0.05-0.04}\pm0.02 $       & $0.19^{+0.07}_{-0.05} $ & $0.20^{+0.08}_{-0.06} $  &$0.3$\; \cite{chen07} & \\
$Br(\bar{B}^0\to\eta^\prime \tau^+ \tau^-)(10^{-9})$ &0.21 & $0.25^{+0.07+0.05+0.02}_{-0.05-0.04-0.01}$     & $0.20^{+0.07}_{-0.05} $ & $0.21^{+0.08}_{-0.06} $  & & \\
$Br(\bar{B}^0\to\eta^\prime \nu\bar\nu)(10^{-9})$    &0.14 & $0.18 ^{+0.05+0.04+0.02}_{-0.04-0.03-0.02}$    & $0.14^{+0.05}_{-0.04} $ & $0.16^{+0.06}_{-0.05} $  & & \\
%%%%%%%%%%%%%
$Br(\bar{B}_s^0\to\eta l^+ l^-)(10^{-7})$            &1.68 & $2.07 ^{+0.65+0.57+0.10}_{-0.51-0.50-0.09}$    & $2.59^{+1.09}_{-0.90} $ & $2.20^{+0.93}_{-0.77}$   &$2.4$\cite{choi10,azizi10}; & \\
$Br(\bar{B}_s^0\to\eta \tau^+ \tau^-)(10^{-7})$      &0.39 & $0.45 ^{+0.15+0.13}_{-0.11-0.11}\pm0.02 $      & $0.56^{+0.25}_{-0.21} $ & $0.48^{+0.21}_{-0.18} $  &$0.34$\cite{wu06}  & \\
$Br(\bar{B}_s^0\to\eta \nu\bar\nu)(10^{-6})$         &1.33 & $ 1.62 ^{+0.54+0.45+0.11}_{-0.38-0.39-0.10}$   & $2.03^{+0.89}_{-0.69} $ & $1.72^{+0.76}_{-0.59} $  &$1.4$\cite{azizi10} &  \\
%%%%%%%%%%%%%
$Br(\bar{B}_s^0\to\eta^\prime l^+ l^-)(10^{-7})$     &1.77 & $ 2.18 ^{+0.73+0.61}_{-0.52-0.53}\pm0.15$      & $1.45^{+0.64}_{-0.50} $ & $1.92^{+0.85}_{-0.67} $  &$1.8$\cite{choi10}& \\
$Br(\bar{B}_s^0\to\eta^\prime \tau^+\tau^-)(10^{-7})$&0.23 & $ 0.27 ^{+0.09}_{-0.07}\pm0.07\pm0.01 $        & $0.18^{+0.07}_{-0.06} $ & $0.24^{+0.10}_{-0.09} $  &$0.28$\cite{azizi10}& \\
$Br(\bar{B}_s^0\to\eta^\prime \nu\bar\nu)(10^{-6})$  &1.39 & $ 1.71^{+0.57+0.47}_{-0.41-0.42}\pm0.12$       & $1.14^{+0.47}_{-0.40} $ & $1.50^{+0.62}_{-0.53} $  &$1.3$\; \cite{azizi10} & \\
\hline \hline
\end{tabular} }
\end{center}
\end{table}

%%%%%%%%%%%%%%%%%%%%%%%%%%%%%%%%%%%%%%%%%%%%%%%%vvv

In Table \ref{tab-Bq-Bss} we list the LO and NLO pQCD predictions for the form factors
$F_{0,+,T}(0)$ involved in the $B\to\eta_q$ and $B_s\to\eta_s$ transitions.
The three errors of $F_{0,+,T}(0)$ come from the uncertainties
of $\omega_B=0.40\pm 0.04$ GeV or $\omega_{B_s}=0.50\pm 0.05$ GeV,
$f_B=0.21\pm0.02$ GeV or $f_{B_s}=0.23\pm0.03$ GeV, and
$a_2^{\eta_q,\eta_s}=0.115 \pm 0.115$, respectively.
The errors from the variations of $V_{ts}$ or $|V_{td}/V_{ts}|$, and $f_q=(1.07\pm0.02)f_\pi$
or $f_s=(1.34\pm0.06)f_\pi$ are very small, and have been neglected.
One can see that the NLO contribution to the form factors $F_{0,+,T}(0)$ in the
$B/B_s \to (\eta_q, \eta_s)$ transitions can provide about $12\%$ enhancement to
the LO ones. The NLO pQCD predictions for $F_{0,+,T}^{B_{(s)}\to\eta_{q,s}}(0)$
agree well with the values obtained from other methods.

The pQCD predictions (in units of $10^{-2}$) for the form factors
$F_{0,+,T}(0)$ in the $B/B_s\to\eta_g$ transitions are also shown in Table \ref{tab-Bq-Bss}.
The sources of the first two errors are the same as those for the $B\to\eta_q$ and $B_s\to\eta_s$
transitions, while the third one comes from the uncertainty of $B_2=4.6\pm2.5$.
It is easy to see that the form factors $F_{0,+,T}(0)$ in the $B/B_s\to\eta_g$ transitions
are of order $10^{-3}$, so the corresponding contributions to the decay rates are
negligible. Using the relevant formulas and the input parameters
given above, it is straightforward to calculate the branching
ratios of the considered decays.

In the FKS mixing scheme, the LO and NLO pQCD predictions for the branching ratios of
the considered decays with $l=(e,\mu)$ are listed in Table \ref{tab-brn1}.
We show only the central values of the LO pQCD predictions in column two, and
the central values and the major theoretical errors simultaneously in column
three. The first error arises from the uncertainty of $\omega_B=0.40\pm 0.04$ or
$\omega_{B_s}=0.50\pm 0.05$, the second one from the uncertainty of
$f_B=0.21\pm0.02$ or $f_{B_s}=0.23\pm0.03$, and the third one is induced by the
variations of $a_2^{\eta_q,\eta_s}=0.115\pm 0.115$. One can see from
Table \ref{tab-brn1} that:
\begin{itemize}
\item
For all the considered decays, the inclusion of the NLO contribution to the
$B_{(s)} \to \eta_{q,s}$ transition form factors provide about $25\%$ enhancement to
the branching ratios. The pQCD predictions for the $B^-\to \eta^{(')} l^-\bar{\nu}_l$ decay
rates then become well consistent with the measured values and other known theoretical predictions
\cite{kim2001,prd-75-054003}.

\item
For all neutral current processes, the NLO pQCD predictions basically agree with
other known theoretical predictions \cite{chen07,choi10,chen10,wu06}.

\item
Because of $Br(\bar{B}_s^0\to\eta^{(\prime)} \nu\bar\nu)\approx 1.7\times 10^{-6}$, these
decays may be observed at the LHCb. Other neutral decay modes with the decay rates at
$10^{-7} - 10^{-9}$ level may be very hard, if not impossible, to measure.

\end{itemize}
%%======================================

\begin{table}[thb]
\begin{center}
\caption{pQCD predictions for the
$B^-\to Gl^-\bar\nu_l(\tau^-\bar\nu_\tau)$
and $\bar{B}^0_{(s)}\to Gl^+l^-(\tau^+\tau^-, \nu\bar\nu)$ branching ratios
in the $\eta$-$\etar$-$G$ mixing scheme for $\phi_G=33^\circ$ or $22^\circ$. }
\label{tab-br-B2G-CLL}
\begin{tabular}{l| c| c}  \hline\hline
 Decay modes
 & \hspace{0.5cm}$\phi_G=33^\circ$  &\hspace{0.5cm}$\phi_G=22^\circ$ \\
\hline %%%%%%%%%%%%%
$Br(B^-\to G l^- \bar\nu_l)(10^{-5})$ \;
&\; $0.64^{+0.23}_{-0.22} $
&\; $0.30^{+0.11}_{-0.10} $ \\
$Br(B^-\to G \tau^- \bar\nu_\tau)(10^{-5})$ \;
&\; $0.25^{+0.09}_{-0.08} $
&\; $0.12\pm0.04 $ \\
\hline %%%%%%%%%%%%%
$Br(\bar{B}^0\to G l^+ l^-)(10^{-9})$ \;
&\; $0.76^{+0.28l}_{-0.23}$
&\; $0.36^{+0.13}_{-0.11} $ \\
$Br(\bar{B}^0\to G \tau^+ \tau^-)(10^{-10})$ \;
&\; $0.18^{+0.07}_{-0.05} $
&\; $0.08^{+0.03}_{-0.02} $ \\
$Br(\bar{B}^0\to G \nu\bar\nu)(10^{-9})$ \;
&\; $5.89^{+2.16}_{-1.74} $
&\; $2.79^{+1.02}_{-0.82} $ \\
\hline  %%%%%%%%%%%%%
$Br(\bar{B}^0_s\to G l^+ l^-)(10^{-7})$ \;
&\; $0.24^{+0.11}_{-0.09} $
&\; $0.11^{+0.05}_{-0.04} $ \\
$Br(\bar{B}^0_s\to G \tau^+ \tau^-)(10^{-9})$ \;
&\; $0.88^{+0.38}_{-0.30} $
&\; $0.42^{+0.18}_{-0.14} $ \\
$Br(\bar{B}^0_s\to G \nu\bar\nu)(10^{-7})$ \;
&\; $1.85^{+0.82}_{-0.65} $
&\; $0.88^{+0.39}_{-0.31} $ \\
\hline \hline
 \end{tabular}
 \end{center}
\end{table}%5555555555555555555555555555555

%---------------------------------------------------------------------------------------------------------------%

In the $\eta$-$\etar$-$G$ mixing scheme the physical states $\eta$, $\etar$
and $G$ are related to the flavor states $\eta_q, \eta_s$ and $\eta_g$ through the
mixing matrix $U(\theta,\phi,\phi_G)$ \cite{li-79-014024}.
The $B/B_s\to\eta_g$ transition form factors are two orders of magnitude smaller than
the $B/B_s \to \eta_q,\eta_s$ ones as indicated in Table I, so
the former contributions to the $B/B_s \to\eta,\etar, G$ decays can be
neglected safely. In Table II we also list the NLO pQCD predictions for the
branching ratios in the $\eta$-$\etar$-$G$ mixing scheme with two sets of
mixing angles $(\phi,\phi_G)$: Set-A with $(\phi,\phi_G)=(43.7^\circ,33^\circ)$~\cite{liu-86-011501},
and Set-B with $(\phi,\phi_G)=(40^\circ,22^\circ)$~\cite{li-79-014024}.
The theoretical errors from the uncertainties of $\omega_B,
\omega_{B_s}$, $f_B, f_{B_s}$, and the Gegenbauer moments are added in quadrature.
The pQCD predictions for the $B_{(s)}\to G(ll,l\nu,\nu\bar\nu)$ branching ratios in the
$\eta$-$\etar$-$G$ mixing scheme are listed in Table~\ref{tab-br-B2G-CLL}. One can see
from Tables II and III that:
\begin{itemize}
\item
The pQCD predictions in the two-state mixing schemes, and in the three-state mixing
scheme with the two sets of mixing angles, are all similar within theoretical errors.

\item
The pQCD predictions for $Br(B^-\to Gl^-\bar\nu_l)$ and
$Br(\bar{B}^0_{(s)}\to Gl^+l^-)$ are about two orders of magnitude smaller than
those of the corresponding $B/B_s\to \eta^{(\prime)}$ transitions
as expected.

\end{itemize}

In summary, we have studied the semileptonic decays
$B/B_s \to (\eta,\etar, G)(l^+l^-,l\bar{\nu},\nu\bar{\nu} )$ in this
work by employing the pQCD factorization approach beyond
LO. Based on the numerical calculations and the
phenomenological analysis, the following points have been observed:
\begin{enumerate}
\item[(i)]
The NLO contributions can
provide $25\%$ enhancement to the LO pQCD predictions for
$Br(B^-\to \eta^{(')} l^-\bar{\nu}_l)$, leading to a good agreement
between the pQCD predictions and the data.

\item[(ii)]
For all the considered decays, the pQCD results are basically consistent with
those from other different theoretical models.

\item[(iii)]
The pQCD predictions in the two considered meson mixing schemes agree well
with each other within theoretical errors.

\end{enumerate}

%---------------------------------------------------------------------------------------------------------------%
\begin{acknowledgments}
The authors are very grateful to Hsiang-nan Li for his help.
This work is supported by the National Natural Science
Foundation of China under the Grant No.~10975074 and~11235005.

\end{acknowledgments}
%---------------------------------------------------------------------------------------------------------------%
%\appendix

%%%----------------------------------------------------------------------------------------
%========================= reference=========================%

%========================= reference=======END===============%


\begin{thebibliography}{99}

\bibitem{lipkin91}
H.J.~Lipkin, \plb {\bf 254}, 247 (1991).
%% Interference effects in Kq and Kq' decay modes of heavy mesons.
%% Clues to understanding weak transitions and CP violation

\bibitem{yy01}
M.Z.~Yang and Y.D.~Yang, \npb {\bf 609}, 469 (2001);
Y.-Y.~Keum, H.N.~Li and A.I.~Sanda, \prd {\bf 63}, 054008 (2001);
C.D.~L\"u, K.~Ukai and M.Z.~Yang, \prd {\bf 63}, 074009 (2001).

\bibitem{bn651}
M.~Beneke and M.~Neubert, \npb {\bf 651}, 225 (2003).

\bibitem{kou02}
E.~Kou and A.~Sanda, \plb {\bf 525}, 240 (2002).

\bibitem{zjx2008}
Z.J.~Xiao, Z.Q.~Zhang, X.~Liu, and L.B.~Guo, \prd {\bf 78}, 114001 (2008).

\bibitem{fan2012}
Y.Y.~Fan, W.F.~Wang and Z.J.~Xiao, \prd {\bf 87}, 094003 (2013).
%% Anatomy of $B \to K \eta^{(\prime)}$ decays in different mixing schemes and
%% effects of NLO contributions in the pQCD approach

\bibitem{kagan}
S.~Khalil and E.~Kou, \prl {\bf 91}, 241602 (2003);
A.L.~Kagan and A.A.~Petrov, hep-ph/9707354.

\bibitem{xiao99}
G.R.~Lu, Z.J. Xiao, H.K. Guo and L.X. L\"u, \jpg  {\bf 25}, L85 (1999);
Z.J.~Xiao, K.T.~Chao and C.S.~Li, \prd {\bf 65}, 114021(2002);
Z.J.~Xiao and W.J.~Zou, \prd {\bf 70}, 094008 (2004).

\bibitem{kim2001}
C.S.~Kim and Y.D.~Yang, \prd {\bf 65}, 017501~(2001).
%% Study of the semileptonic decays B^\pm \to \eta^{(\prime)} l \nu

\bibitem{kim-plb590-223}
C.S.~Kim, S.~Oh and C.~Yu, \plb {\bf 590}, 223~(2004).
%% B ¡ú¦Ç^{(\prime)} l¦Í decays and the flavor-singlet form factors

\bibitem{prd-75-054003}
A.G.~Akeroyd, C.H.~Chen and C.Q.~Geng \prd {\bf 75},054003 (2007).
%% $B\to\eta^{(')}(l^-\bar{\nu}_l,l^+l^-,K,K*})$ decays in the quark-flavor mixing scheme

\bibitem{chen07}
C.H.~Chen and C.Q.~Geng, \plb {\bf 645}, 197~(2007).
%% ¦Ç^{(\prime)} productions in semileptonic B decays

\bibitem{choi10}
H.M.~Choi, \jpg {\bf 37}, 085005 (2010).
%% Exclusive rare Bs ¡ú (K, ¦Ç, ¦Ç')l^+l^- decays in the light-front quark model

\bibitem{chen10}
C.H.~Chen, Y.L.~Shen and W.~Wang, \plb {\bf 686}, 118~(2010).
%% |Vub| and B ¡ú¦Ç(') form factors in covariant light-front approach

\bibitem{wu06}
Y.L.~Wu, M.~Zhong and Y.B.~Zuo, \ijmpa {\bf 21}, 6125 (2006). 	
%% B(s), D(s) ---> pi, K, eta, rho, K*, omega, phi Transition FFs and Decay Rates
%%  with Extraction of the CKM parameters |V(ub)|, |V(cs)|, |V(cd)| hep-ph/0604007

\bibitem{azizi10}
K.~Azizi, R.~Khosravi and  F.~Falahati, \prd {\bf 82}, 116001 (2010).
%% Rare semileptonic Bs decays to \eta and \etap mesons in QCD

\bibitem{semil-2012a}
Sergi Gonzalez-Sols, Proceedings of 16th Frascati Spring
School ¡®Bruno Touschek¡¯ in Nuclear Subnuclear and
Astroparticle Physics $\&$ 3rd Young Researchers Workshop on Physics 
Challenges in the LHC Era, Frascati, Rome, Italy, 2012, econf.C12-05-07.1, 49 (2012).
%% eta - eta-prime mixing angle from the SL decays D+ ---> eta(eta') e+ nu/e, 
%%  D/s+ ---> e+ nu/e and B+ ---> eta(eta') l+ nu/l.

\bibitem{pdg2012}
J.~Beringer {\it et al.} (Particle Data Group),  \prd {\bf 86}, 010001 (2012).

%%========================================================

\bibitem{liu-86-011501}
X.~Liu, H.N.~Li and Z.J.~Xiao, \prd {\bf 86}, 011501(R) (2012).
%% Implications on $\eta-eat'-$glueball mixing from $B_{d/s}\to J/\psi\eta'$ Decays

\bibitem{li2003}
H.N.~Li, \ppnp {\bf 51}, 85 (2003); and references therein.

\bibitem{babar-83-052011}
P.~del~Amo~Sanchez et al. (BABAR Collaboration), \prd {\bf 83}, 052011 (2011).
%% Meas. of the B0\to \pi l\nu and B\to \eta l\nu bfs, the B\to \pi l\nu and
%% ... form-factor shapes, and determination of |Vub|
J.P. Lees et al. (BABAR Collaboration), \prd {\bf 86}, 092004 (2012).

\bibitem{prd86-114025}
W.F.~Wang and Z.J.~Xiao, \prd {\bf 86}, 114025 (2012).

\bibitem{li-85074004}
H.N.~Li, Y.L.~Shen and Y.M.~Wang, \prd {\bf 85}, 074004 (2012).
%% NLO corrections to B \to \pi form factors in kT factorization

\bibitem{fks98}
T.~Feldmann, P.~Kroll, and B.~Stech, \prd {\bf 58}, 114006 (1998);
\plb {\bf 449}, 339 (1999);
T.~Feldmann, {\bf Int. J. Mod. Phys. A} {\bf 15}, 159 (2000).
%% Mixing and decay constants of pseudoscalar mesons
%% Mixing and decay constants of pseudoscalar mesons: the sequel
%% Quark structure of pseudoscalar mesons

\bibitem{li-79-014024}
H.Y.~Cheng, H.N.~Li, and K.F.~Liu, \prd {\bf 79}, 014024 (2009).
%% Pseudoscalar glueball mass from $\eta-\eta'-G$ mixing

\bibitem{pball-71-014015}
P.~Ball and R.~Zwicky, \prd {\bf 71}, 014015 (2005).
%% New results on B \to \pi, K, \eta decay form factors from light-cone sum rules

\bibitem{wfun-pball-pi}
P.~Ball, \jhep {\bf 01}, 010 (1999).
%% TH Update of Pseudo-S-Meson DAs of Higher Twist: The Nonsinglet Case

\bibitem{li-74-074024}
Y.Y.~Charng, T.~Kurimoto and H.N.~Li, \prd {\bf74}, 074024 (2006) ;
\prd {\bf78}, 059901(E) (2008).
%% Gluonic contribution to $B\to\eta^{(\prime)}$ from factors

\bibitem{ali-367}
A.~Ali and A.Y.~Parkhomenko, \epjc {\bf 30}, 183 (2003);
%% The \eta^\prime g^*g^{(*)} vertex including the \eta^\prime-meson mass
\epjc {\bf 30}, 367 (2003).
%% An analysis of the inclusive decay \Upsilon(1S)\to \eta^\prime X and
%%  constraints on the \eta^\prime-meson distribution amplitudes


\bibitem{kroll-67}
P.~Kroll and K.~Passek-Kumericki, \prd {\bf 67}, 054017 (2003).
%% Two-gluon components of the \eta and \eta^\prime mesons
%%  to leading-twist accuracy

\bibitem{npb592-3}
M.~Beneke, T.~Feldmann, \npb {\bf 592}, 3 (2001).
%% Symmetry-breaking corr. to heavy-to-light B meson form factors at large
%%   recoil;


\bibitem{lu-jpg37}
W.~Wang, Y.L.~Shen and C.D.~L\"{u}, \jpg {\bf 37}, 085006 (2010).
%% B-to-Glueball form factor and Glueball production in B decays

\end{thebibliography}
\end{document}